A15 Nb$_3$Si: A "high" T$_c$ superconductor synthesized at a pressure of one megabar and metastable at ambient conditions

Jinhyuk Lim, J. S. Kim, Ajinkya C. Hire, Yundi Quan, R. G. Hennig, P. J. Hirschfeld, J. J. Hamlin, and G. R. Stewart

Department of Physics, University of Florida, Gainesville, FL 32611

Bart Olinger

Los Alamos National Laboratory, Los Alamos, NM 87545

**Abstract:** A15 Nb$_3$Si is, until now, the only "high" temperature superconductor produced at high pressure (~110 GPa) that has been successfully brought back to room pressure conditions in a metastable condition. Based on the current great interest in trying to create metastable-at-room-pressure high temperature superconductors produced at high pressure, we have restudied explosively compressed A15 Nb$_3$Si and its production from tetragonal Nb$_3$Si. First, diamond anvil cell pressure measurements up to 88 GPa were performed on explosively compressed A15 Nb$_3$Si material to trace T$_c$ as a function of pressure. T$_c$ is suppressed to ~ 5.2 K at 88 GPa. Then, using these T$_c$ (P) data for A15 Nb$_3$Si, pressures up to 92 GPa were applied at room temperature (which increased to 120 GPa at 5 K) on tetragonal Nb$_3$Si. Measurements of the resistivity gave no indication of any A15 structure production, i. e. no indications of the superconductivity characteristic of A15 Nb$_3$Si. This is in contrast to the explosive compression (up to P~110 GPa) of tetragonal Nb$_3$Si, which produced 50-70 % A15 material, T$_c$ = 18 K at ambient pressure, in a 1981 Los Alamos National Laboratory experiment. This implies that the accompanying high temperature (1000 ºC) caused by explosive compression is necessary to successfully drive the reaction kinetics of the tetragonal → A15 Nb$_3$Si structural transformation. Our theoretical calculations show that A15 Nb$_3$Si has an enthalpy vs the tetragonal structure that is 70 meV/atom *smaller* at 100 GPa, while at ambient pressure the tetragonal phase enthalpy is lower than that of the A15 phase by 90 meV/atom. The fact that "annealing" the A15 explosively compressed material at room temperature for 39 years has no effect shows that slow kinetics can stabilize high pressure metastable phases at ambient conditions over long times even for large driving forces of 90 meV/atom.

I.     Introduction

In the past several years, the discoveries[1-4] of superconductivity in various hydrides at temperatures above 200K and pressures above 1 megabar have underscored the lack of fundamental limits on electron-phonon superconducting critical temperatures. Further, these breakthrough experiments have raised the question: can a way be found to stabilize these high temperature superconductors back down to ambient pressure in order to achieve room-temperature applications?

One possible approach would be the creation of new metastable superconducting structures at high pressures which survive decompression due to a thermodynamic barrier blocking transition back to the ground state. The present work studies an outstanding example of metastable superconductivity in $Nb_3Si$, focusing on the structures before and after compression, and the thermal stability of the metastable high pressure A15 phase at ambient pressure.

The A15 superconductors of general form $Nb_3X$, with X=Al, Ga, Sn, and Ge were the highest temperature superconductors known[5] starting in the 1950s until the discovery of the cuprate superconductors in 1986. The highest $T_c$ in this family occurs at ambient pressure in $Nb_3Ge$, discovered in 1973.

In 1981 Olinger and Newkirk[6] succeeded in converting tetragonal $Nb_3Si$ into the (3% smaller unit cell) cubic A15 structure via explosive compression using plastic explosive (RDX plus other minor ingredients). Characterization via x-ray diffraction[6] and bulk low temperature specific heat measurements[7] indicated a 50-70% conversion of the tetragonal phase into the superconducting A15 phase, with a $T_c$ onset of 18.0 K. Hydrodynamic calculations[6] indicated that the tetragonal phase sample experienced a pressure ~90-110 GPa, with a peak temperature of ~1000 °C, a residual temperature of ~ 500 °C, and a time[8] at peak pressure of ~1 μsec.

As part of our program of producing new superconductors under high pressures using diamond anvil techniques, we used some of the same tetragonal starting $Nb_3Si$ material as used at Los Alamos National Laboratory. This tetragonal material was subjected in the present work to pressures up to 120 GPa in a diamond anvil cell to search for superconductivity from any converted A15 phase. Based on explosive compression treatment of another A15 material, $V_3Si$, a sample of A15 $Nb_3Si$ converted more 'gradually' from the tetragonal structure could have significantly improved superconducting properties, e. g. $T_c$ of explosively compressed A15 $V_3Si$ is suppressed[9] vs material that was arc-melted followed by annealing by ~10%.

According to our enthalpy calculations of various $Nb_3Si$ structures, the A15 structure becomes more stable than the tetragonal structure above 50 GPa, with the enthalpy difference growing to about 70 meV/atom (or approximately 550 °C of thermal energy) by 100 GPa. Of course, the elevated temperatures (~ 1000 °C) during the explosive compression preparation technique is likely to have added a necessary ingredient for the observed tetragonal → A15 structure conversion at high pressure. For example, in another case of a compound ($Nb_3Bi$) converted to the A15 structure under an applied pressure (only 3 GPa in this instance), the kinetics of formation for the A15 structure[10] required the application of 1000 °C.

At ambient pressure, we calculate that the tetragonal structure of $Nb_3Si$ is more stable than the A15 structure by 90 meV/atom. Since the sample of A15 $Nb_3Si$ produced by explosive compression at Los Alamos in 1981 that was studied in the present work was unchanged after storage at room temperature for 39 years, it is thus clear that slow kinetics can stabilize high-pressure metastable phases at ambient conditions even for large driving forces of 90 meV/atom.

The only $T_c$ vs P data of A15 Nb3Si are from Lim et al.[11] up to 2 GPa (dTc/dP = -0.267 K/GPa). In order to know what $T_c$ to search for in our up-to-120 GPa attempt at conversion, we further characterized some of the original explosively compressed material ("recovery 20") from the 1981 Los Alamos experiments[6]. This characterization has two goals. The first is to determine the $T_c$ of the A15 $Nb_3Si$ structure at these high pressures. In the tetragonal phase starting material, the presence of the minority Nb second phase introduces a $T_c$ ~8-9 K (relatively[12] pressure independent, $\Delta T_c$ ~ -2 K at 100 GPa.) If the P≤2 GPa data of Lim et al.[11] are linearly extrapolated, this would give $T_c$ (A15 $Nb_3Si$) crossing with $T_c$ of the Nb second phase at around 40 GPa. This would imply that the second phase Nb $T_c$ would need to be separated from any converted A15 signal at higher pressures. The second reason to further characterize the existing A15 $Nb_3Si$ was to conduct annealing experiments to determine the temperature and rate for back conversion (A15 → tetragonal). This will give, as a rough approximation, an estimate of whether pressurizing tetragonal $Nb_3Si$ material up to 92 GPa at only room temperature (which increased to 120 GPa at 5 K) can overcome the inherent sluggishness of the thermodynamic transformation involved.

In Section II we present characterizations of the unusual, explosively compressed A15 $Nb_3Si$ as well as of tetragonal $Nb_3Si$. We then present results in Section III for the resistive transition of the A15 samples under pressure up to 88 GPa. We give details of theoretical structure relaxation calculations suggesting that a transition from tetragonal to A15 phase might take place at pressures of the order of 50 GPa. However, no such transition is observed in our measurements on tetragonal $Nb_3Si$ at pressures up to 120 GPa. To investigate the size of the possible thermal barrier preventing the structural transition, we perform annealing studies of the A15 sample, and find that temperatures of order 600 ºC on timescales of days are sufficient to convert large fractions of the A15 phase to the tetragonal phase, consistent with theoretical enthalpy difference estimates. We characterize the A15 sample at ambient pressure with specific heat and diamagnetic susceptibility measurements before and after annealing. Finally, in Section IV we discuss the relationship of our measurements to the general program of creating useful metastable high-$T_c$ materials.

## II.     Experimental

### a. Samples

Two samples were used in the current study. The first was $Nb_3Si$ (from the original work[1] at Los Alamos National Laboratory) prepared by arc-melting and annealed at approximately 1780 ºC for four hours. As evident from the binary phase diagram[13] for Nb-Si, the desired tetragonal phase of $Nb_3Si$ is stable between 1673 and 1977 °C. According to x-ray diffraction, the annealed sample used in the present work was >80% tetragonal $Nb_3Si$, with bcc Nb and $Nb_5Si_3$ as minority second phases. In order to better quantitatively compare the amount of second phase Nb in this starting material with the explosively compressed Recovery 20 sample discussed below, specific heat at low temperatures was performed.

The second sample used in this study was a piece of the original Recovery 20, explosively compressed $Nb_3Si$ from the work[6] by Olinger and Newkirk which had been stored at room temperature for 39 years. The specific heat of this original, approximately 50-70 % A15, sample was measured for the present work. Then the sample was annealed at increasing temperatures, 1 day at each temperature, starting at 220 °C, with the decreasing amount of A15 material being estimated by dc susceptibility. Upon annealing at one day at 600 °C, the rate of decrease of A15 content was judged to have become large enough to switch to increasing the annealing time (for a total of 5 days) and keeping the annealing temperature fixed at 600 °C. The specific heat of the annealed sample was then measured (Section IIId).

b. **Experimental Measurements**

In addition to performing x-ray diffraction on the starting tetragonal and explosively compressed, A15 majority phase, samples, the following measurements were done. Both these samples were measured resistively in a diamond anvil cell apparatus described below, the tetragonal sample up to 120 GPa to search for possible conversion to the A15 phase and the Recovery 20 sample up to 88 GPa to determine the $T_c$ vs pressure behavior of A15 $Nb_3Si$. The specific heat was measured of 1.) the tetragonal $Nb_3Si$ sample, of 2.) the explosively compressed Recovery 20 A15 $Nb_3Si$ sample, and of 3.) the annealed (for a total of 5 days at 600 °C, in addition to 1 day at 220, 300, 400 and 500 °C) Recovery 20 sample to determine the amount of the A15 phase remaining after annealing.

The low temperature measurement of specific heat utilized the time constant method, and followed the techniques described in ref. 14.

For the high-pressure resistivity measurements, a micron-sized $Nb_3Si$ (either tetragonal or majority A15) polycrystal sample (~30 × 40 × 10 $\mu m^3$) was cut from a larger piece of bulk sample and placed in a gas-membrane-driven diamond anvil cell (OmniDAC from Almax-EasyLab) along with a ruby (~10 $\mu$m in diameter) for pressure calibration[15]. Two opposing diamond anvils (~0.14 carat, type Ia) were used, which have 0.35 mm diameter culets beveled at 8° to 0.18 mm central flats. A Re metal gasket (~3 × 4 × 0.25 $mm^3$) was pre-indented to ~25 $\mu$m in thickness and electrospark-drilled to have a hole (~170 $\mu$m in diameter) at the center. The hole was then filled with 4:1 cBN-epoxy mixture to insulate the gasket and mechanically drilled to have a hole (~80 $\mu$m in diameter) at the center, which was filled with soapstone (steatite) for electrically insulating the sample from the gasket and also serving as the pressure-transmitting medium. The thin $Nb_3Si$ sample was then placed on top of four thin (4 $\mu$m) Pt leads embedded in the insulation layer for a four-point dc electrical resistivity measurement. Further details of the nonhydrostatic high-pressure resistivity technique are given in a paper by Shimizu *et al.*[16]

The diamond cell was placed inside a customized continuous-flow cryostat (Oxford Instruments). A homebuilt optical system attached to the bottom of the cryostat was used for the visual observation of the sample and for the measurement of the ruby manometer. Pressure was applied (or unloaded) at room temperature (~297 K) to the desired pressure which then increased upon cooling, e. g. from 92 GPa at room temperature to 120 GPa at 5 K. Then the sample was cooled down to ~3 K and warmed up again to room temperature at a rate of ~0.5 K/min at each pressure for the temperature-dependent resistivity measurement. Pressures used in the $T_c$ vs pressure phase diagram (Fig. 1 below) were measured at around 5 to 20 K, in which range pressure is constant to within 1 GPa, and the estimated accuracy is ±5%. To estimate the electrical resistivity from the resistance, we used the van der Pauw method, (assuming an isotropic sample in the measurement plane), $\rho = \pi t R/\ln 2$, where $t$ is the sample thickness (~10 $\mu$m) with a current of 1 mA. The accuracy of the estimated resistivity is roughly a factor of two considering uncertainties in the initial thickness of the sample. No attempt was made to take into account the changes in the sample thickness under high pressures.

### III. Results and Discussion

#### a.) High pressure determination of $T_c$ vs pressure in A15 $Nb_3Si$.

In order to better understand the results of our high pressure search for A15 conversion of tetragonal $Nb_3Si$ up to 120 GPa, we discuss our $T_c$ vs pressure results on Recovery 20, A15 $Nb_3Si$ first. As shown in Fig. 1, starting at 2 GPa $T_c$ of A15 $Nb_3Si$ falls monotonically at a rate of about 0.2 K/GPa, consistent with the earlier work of Lim et al.[11] Above about 40 GPa two things happen in the $T_c$ vs pressure data on A15 $Nb_3Si$ in Fig. 1. First, the rate of decrease in $T_c$ with increasing pressure starts to level off. Second, the presence of ~15 % second phase of Nb in the explosively compressed material (see specific heat data discussed below) shows up at ≥ 55 GPa as a second resistive transition (red squares) *above* the A15 transition. Since the weak pressure dependence of this upper transition above 40GPa is consistent with that[12] of pure Nb under pressure, we conclude that the ~15% Nb volume fraction determines the onset Tc in this pressure range. Within this interpretation, the A15 $T_c$ continues to be suppressed above 40 GPa, appearing as a second, lower T, resistive anomaly as seen in Fig. 1.

Representative resistance data at low temperature (during the loading condition) from which the $T_c$ values in Fig. 1 are derived are shown in Fig. 2. The higher temperature transition for P≥55 GPa, as discussed, is identified as being that of the Nb impurity phase. This is consistent with several features of the data. 1.) It is known that $T_c$ of Nb vs pressure is relatively flat[12], which argues for the upper transition at 55 – 88 GPa being that of Nb. 2.) The drop in resistance at the upper transition for 55-88 GPa is only partial, whereas the resistance at the lower transition goes all the way to zero. This implies that the lower transition is the $T_c$ of the majority, 50-70%, A15 phase in the Recovery 20 sample, while the upper, partial resistive transition is the superconducting transition of the minority, 15% (see specific heat data below) Nb impurity phase.

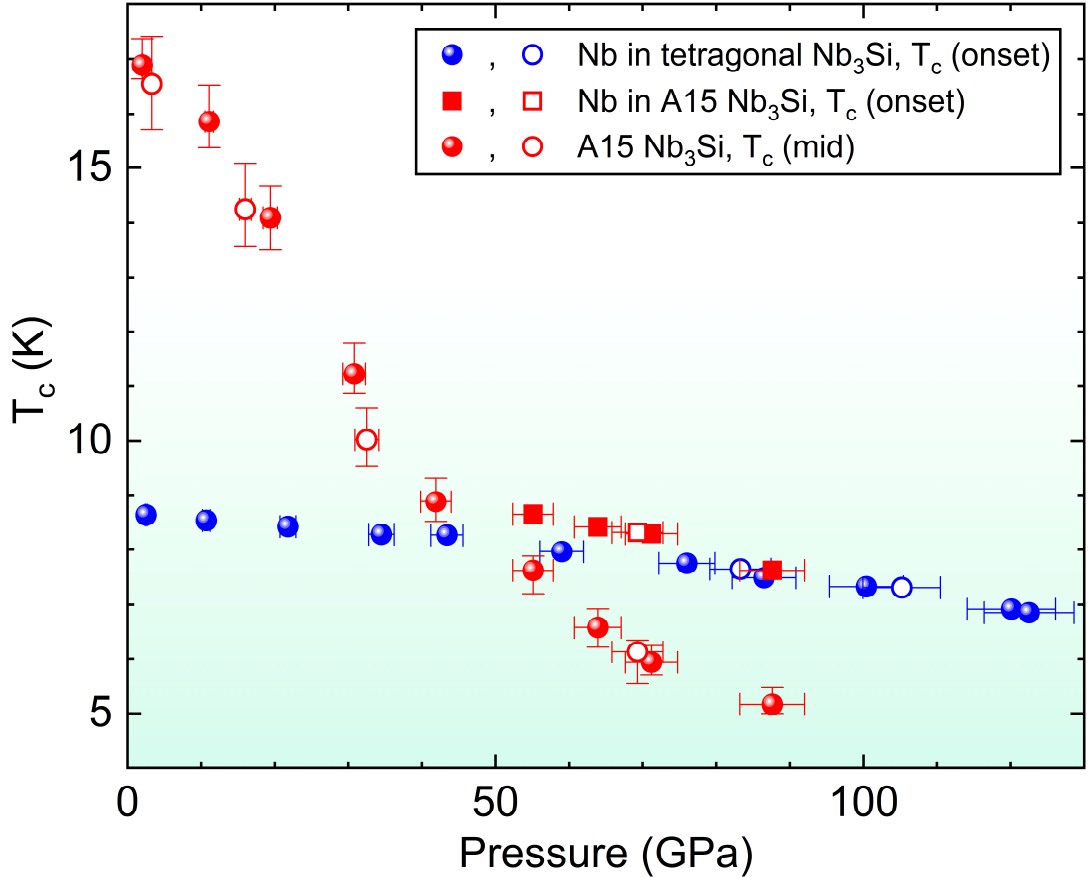

Fig. 1 $T_c$ versus pressure for two experimental runs, one beginning with tetragonal $Nb_3Si$ (blue symbols) and the other with A15 $Nb_3Si$ (red symbols). Unloading Tc values are the open symbols. Unloading the applied pressure on the A15 sample, where the highest pressure applied was only 88 GPa, was successful down to low pressures, as shown by the open red circles. Upon unloading the pressure applied to the tetragonal sample from the highest pressure of 120 GPa (blue circles), below 70 GPa the gasket failed resulting in a sudden release of pressure. In our experimental set up, this is not an unusual occurrence while unloading the pressure from such a high value.

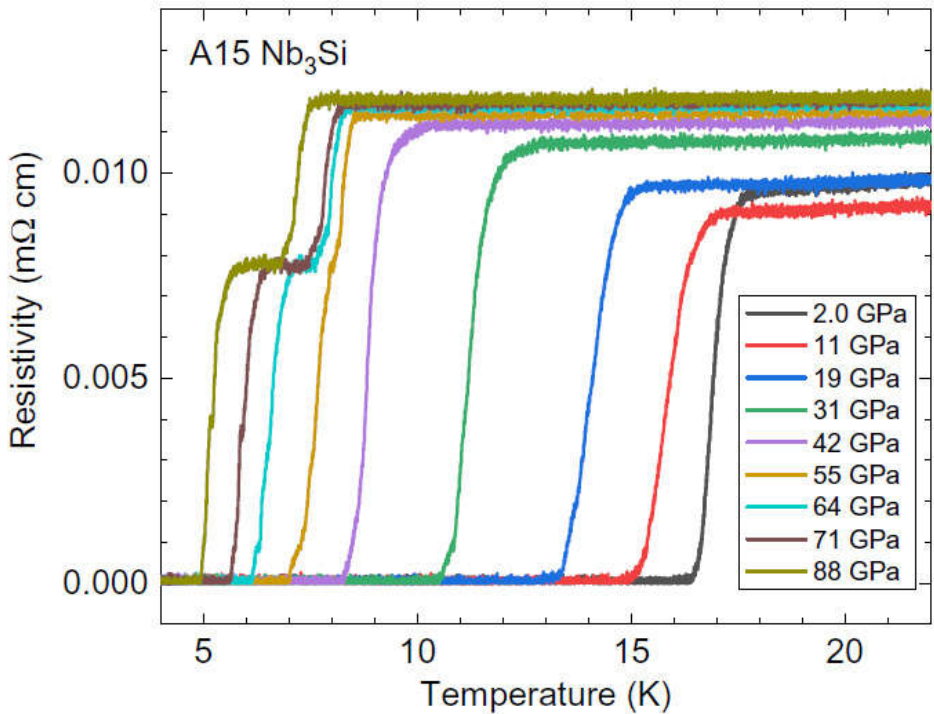

Fig. 2 As can be seen in Fig. 1, where there are two transitions starting at 55 GPa, the resistivity data (while loading) displayed here show how a second transition appears for P≥55 GPa. See text for discussion.

b.) **High pressure resistivity measurement to search for possible tetragonal → A15 conversion up to 120 GPa**

Now that we have established the behavior of $T_c$ vs pressure for A15 $Nb_3Si$, let us consider our high pressure resistance measurements of the tetragonal $Nb_3Si$ sample. Is there any evidence of a $T_c$ from a partial conversion to the A15 structure from applying pressure *at room temperature* to the tetragonal $Nb_3Si$ structure? $T_c$ as determined from resistance measurements on the tetragonal $Nb_3Si$ sample are shown as blue circles in Fig. 1. Due to the ~ 10% second phase of Nb in this tetragonal sample (as determined by specific heat of the starting material as discussed below), there is the constant presence of a partial superconducting transition from Nb starting at around 8.8 K at 2 GPa and going down to ~ 7 K ($T_c^{onset}$) at 120 GPa. Again, this is consistent[12] with measurements on pure Nb with some trace impurities (in our case dissolved Si). In order to settle this question, we present high pressure unloading data on the tetragonal $Nb_3Si$ sample in Fig. 3, where the inserted arrows correspond to the measured data up to 88 GPa (and extrapolated thereafter) for $T_c$ on the explosively compressed ("Recovery 20") $Nb_3Si$ shown in Fig. 1. As can be clearly seen, no apparent anomaly corresponding to the A15 $Nb_3Si$ $T_c$ is observed in our results for high pressure applied to tetragonal $Nb_3Si$ at room temperature. This leaves us with the surmise that the temperatures reached[6] in the explosive compression production (~1000 °C) of A15 $Nb_3Si$ are necessary to overcome thermodynamic sluggishness from the kinetics of the phase transformation.

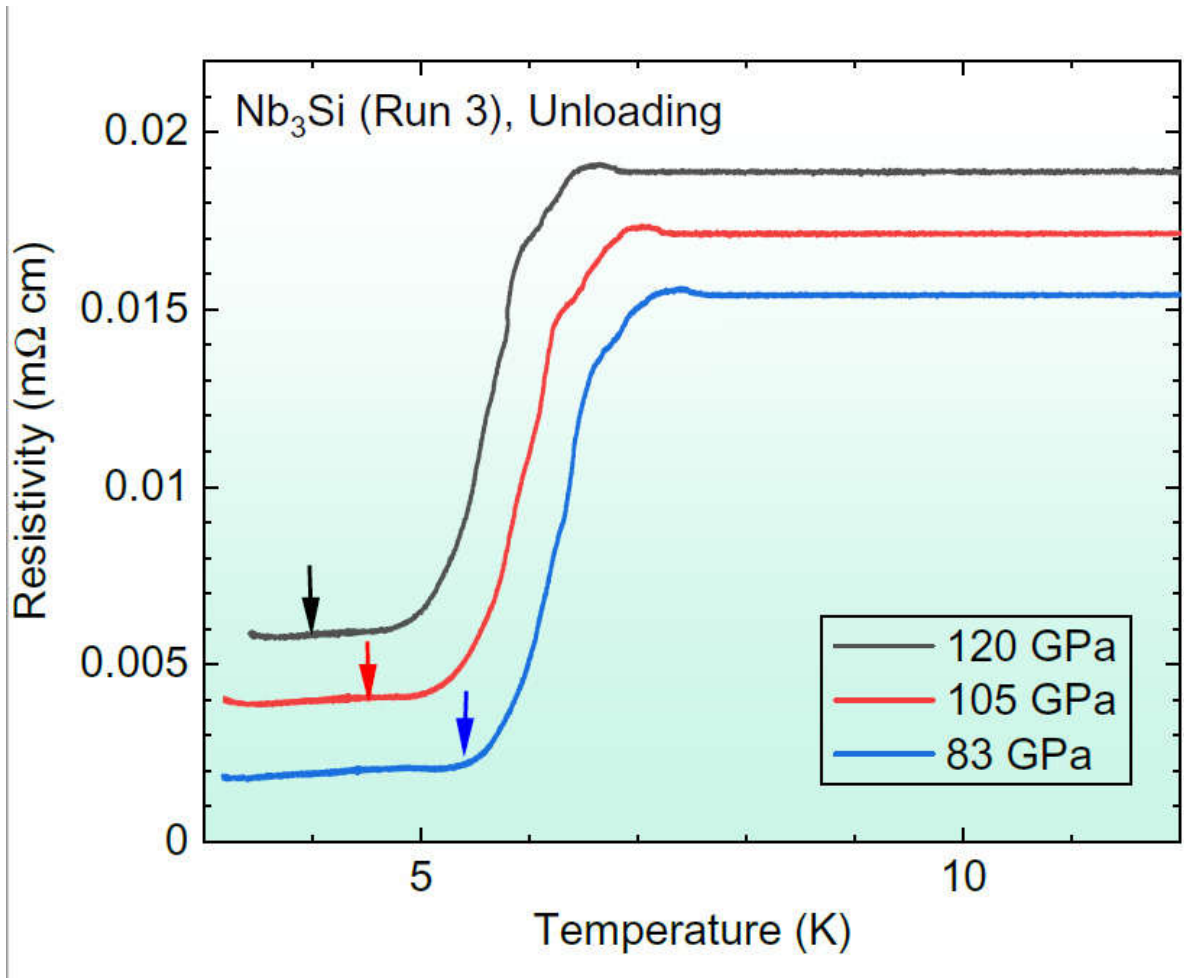

Fig. 3 Resistivity of the tetragonal $Nb_3Si$ sample upon unloading the pressure from 120 GPa (measured at low temperature as discussed in the text). As discussed below, the percentage of Nb second phase in this sample is 10%. Arrows show $T_c$ for A15 extrapolated from data in Fig. 1, and do not correspond to any observable feature in the data shown here.

We now address this issue by performing enthalpy calculations on various possible $Nb_3Si$ structures. We follow these theoretical calculations with annealing experiments on a piece of Recovery 20 A15 majority phase $Nb_3Si$, with the assumption that the thermal conversion A15→non-superconducting tetragonal $Nb_3Si$ gives an approximation of the required thermal energy for the reverse process.

### c.) Enthalpy calculations on Nb$_3$Si

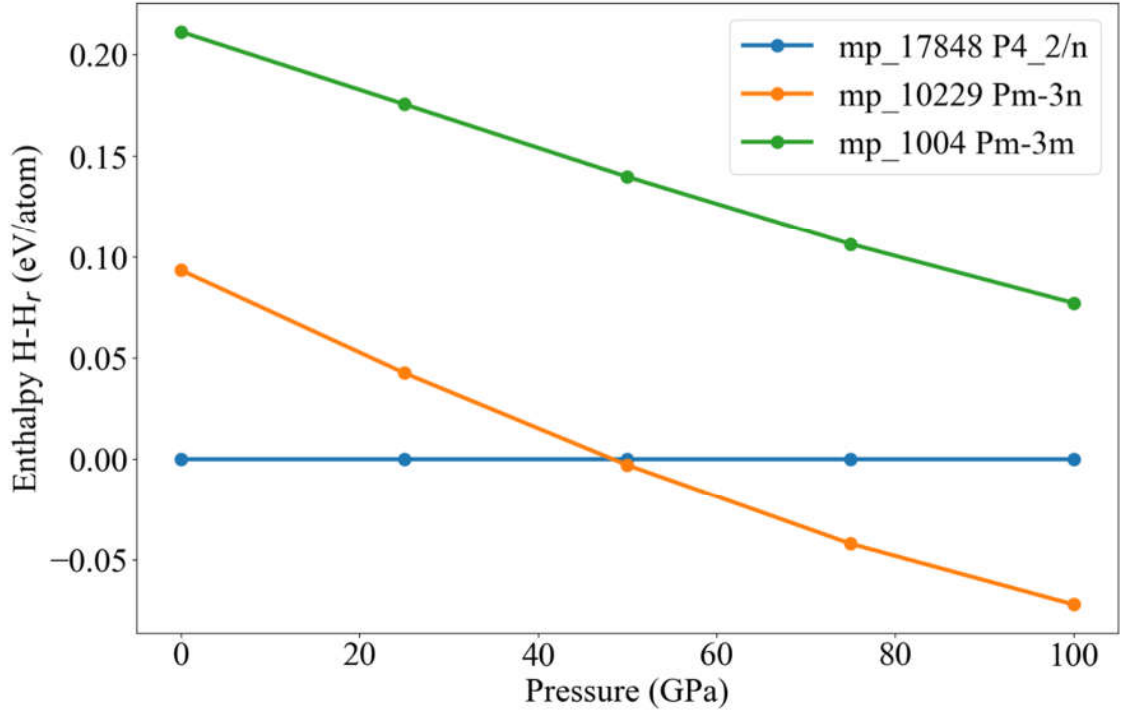

Fig. 4: Enthalpy vs pressure for possible Nb$_3$Si structures. Blue data are for the tetragonal Nb$_3$Si, Materials Project code mp17848 – the starting material where the tetragonal structure is stable between 1673 and 1977 °C - used in the production by explosive compression of the A15 phase in the 1981 Los Alamos National Laboratory experiments. This same material was the one taken to 120 GPa (measured at low temperature) in a diamond anvil apparatus in the present work, Figs. 1 and 3. The orange data represent the enthalpy vs pressure for A15 Nb$_3$Si. The green data represent calculations for the cubic perovskite phase of Nb$_3$Si.

Figure 4 shows our calculation of the enthalpy of Nb$_3$Si as a function of pressure for the three experimentally known crystal structures - $P4_2n$ (tetragonal), $Pm\underline{3}n$ (A15), and $Pm\underline{\ }3\,m$ (cubic perovskite)[17] using density functional theory as implemented in VASP [18-19]. The plane wave cutoff was set to 520 eV and a k-point density of 60 points per Å$^{-1}$ was used for the density functional theory relaxations at varying pressure. We used the Perdew-Burke-Ernzerhof generalized gradient approximation[20] for the exchange-correlation energy and the projector augmented wave pseudopotentials[21]. As shown in Fig. 4, the A15 phase becomes stable above 50 GPa, and at 100 GPa displays an approximately 0.07 eV/atom lower enthalpy than the tetragonal $P4_2n$ phase.

### d.) Annealing experiments on explosively compressed A15 Nb$_3$Si

Without any knowledge of annealing effects on explosively compressed A15 $Nb_3Si$ (other than the fact that it had not reconverted to tetragonal $Nb_3Si$ after being stored at room temperature for 39 years), we began this investigation with a rather low annealing temperature (220 °C) for one day. We discovered that the amount of A15 material present in the 11 mg piece of Recovery 20 which we committed to this annealing was fairly robust against annealing. We successively annealed the sample at ever increasing temperatures (320, 405, 500, and 600 °C) for one day at each temperature, followed by a 4 day anneal at 600 °C. We measured the diamagnetic susceptibility using a dc magnetometer after each anneal to approximate the decrease in the quantity of A15 superconductor present. To avoid interference from the diamagnetic susceptibility contribution from the 15 % Nb second phase (see specific heat determination below), we focused on the magnitude of the diamagnetic signal at 9.5 K, i. e. above the $T_c$ of the second phase Nb.

After 4 day annealing at 600 °C, we stopped the annealing procedure and measured the specific heat of the annealed 11 mg of the explosively compressed, 50-70% A15 sample ("Recovery 20"), and compared these annealed data with the specific heat of the unannealed Recovery 20 sample. These data are shown in Fig. 5, and indicate that 35% of the A15 portion of the sample remains after annealing. The thermal energy needed to aid the reaction kinetics of the conversion of the tetragonal structure of $Nb_3Si$ into the (3% denser) A15 structure is not the same as for the reverse (A15 → tetragonal). However, at least the rather long time at 600 °C needed to partially reconvert the A15 $Nb_3Si$ into the tetragonal structure is clearly consistent with the result that our attempt to convert the tetragonal $Nb_3Si$ to the A15 structure at pressures up to 92 GPa at room temperature (120 GPa at low temperature) did not achieve the required reaction kinetics.

The specific heat data around 9 K are expanded in Fig. 6 to indicate the Nb fraction in the starting, unannealed sample of Recovery 20 (16%) and in the annealed sample (5 %). The specific heat of the starting sample of tetragonal (≥ 80 %) $Nb_3Si$ (see Fig. 1, blue points, and Fig. 3 for resistivity data thereon), not shown, indicates that 10% of that sample is Nb.

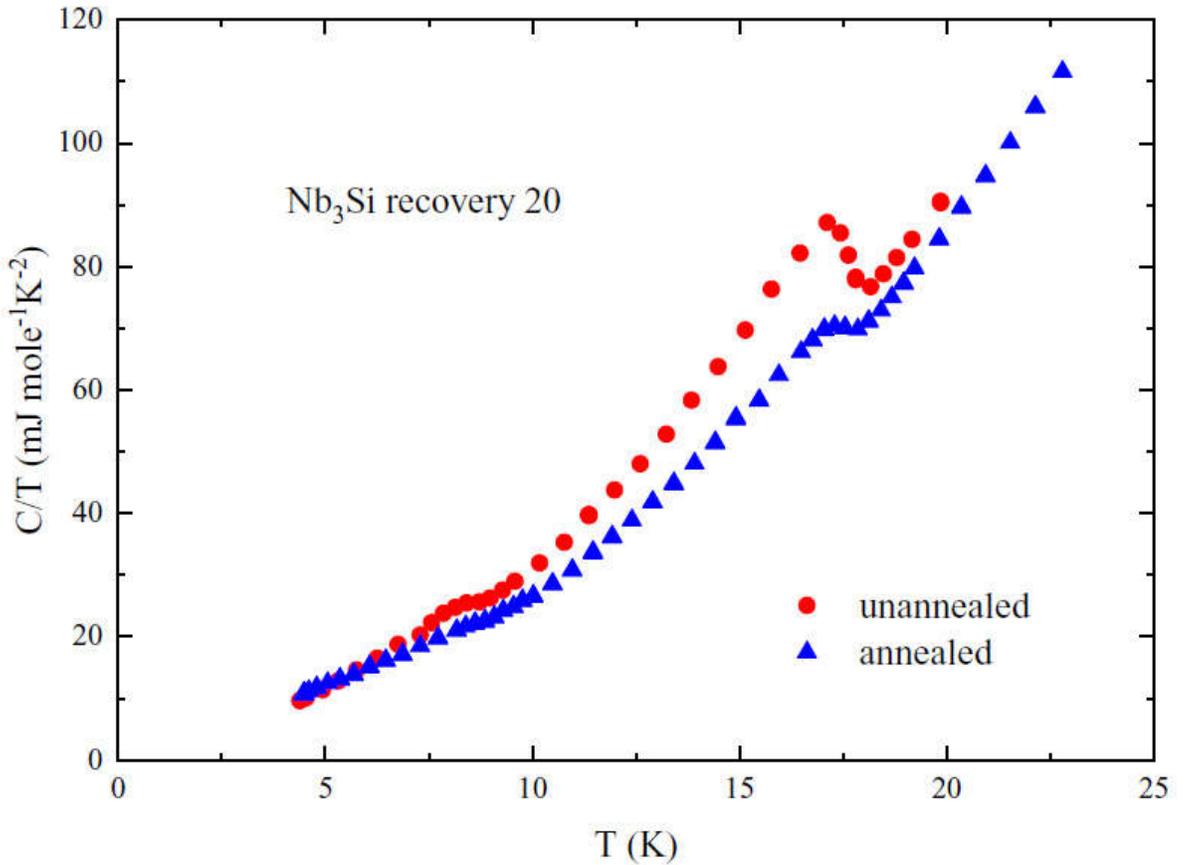

Fig. 5  Shown are data for unannealed and annealed samples of Recovery 20, 50-70 % A15 phase present after explosive compression[6] converted the tetragonal starting material into primarily A15 superconducting phase, $T_c$=18 K.  The unannealed piece had a mass of 32.85 mg, the mass of the piece broken off of this starting piece for annealing was 11.18 mg.  The specific heat of the smaller piece used for annealing showed the same specific heat, normalized per mole, as the larger starting piece, within our error bar of ± 5 %. The annealed piece was subjected successively to one day each at 220, 320, 405, and 500 °C under vacuum, followed by 5 days at 600 °C.  The size of the anomaly in the C/T data at $T_c$, $\Delta C/T_c$, in the annealed sample (blue triangles) is 35 % of the anomaly in the unannealed sample (red circles), both normalized per mole of material.

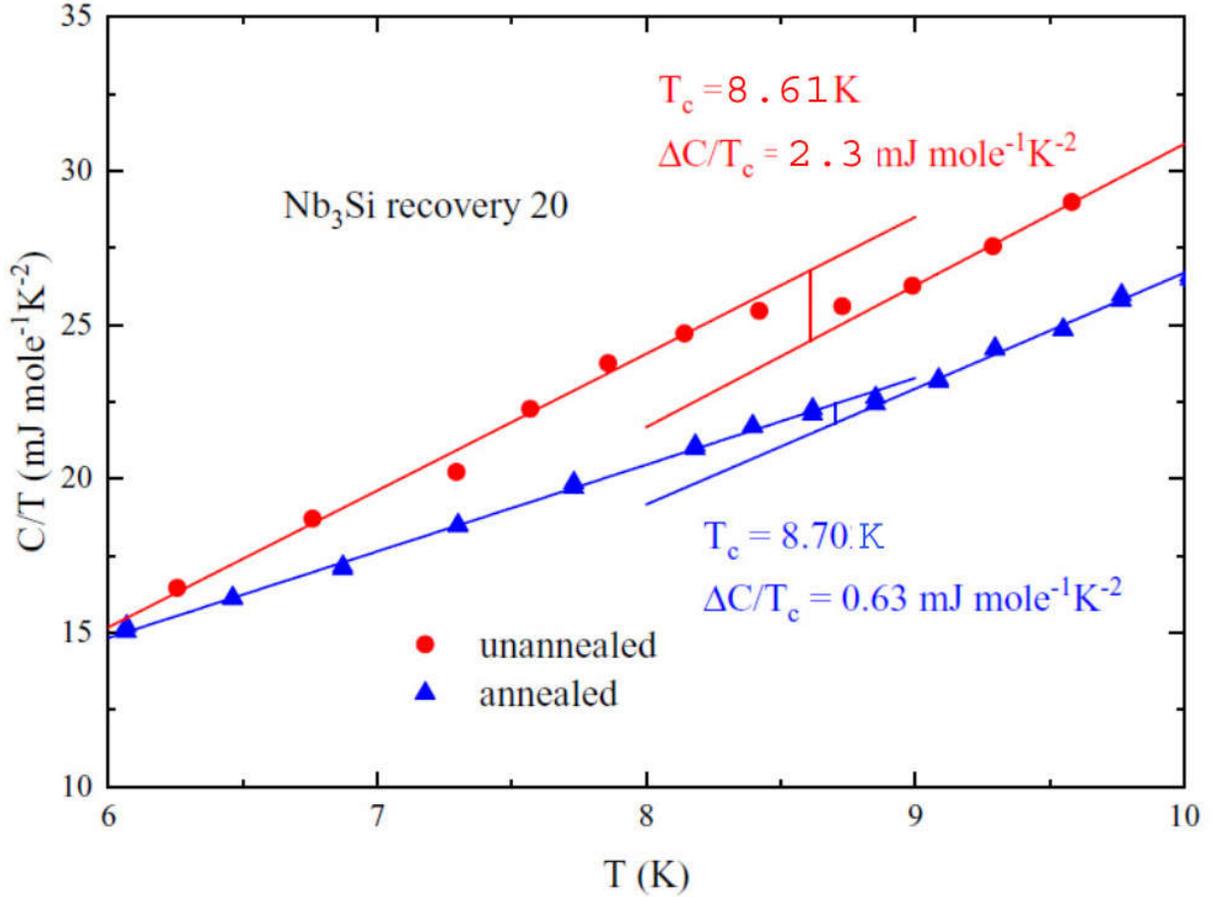

Fig. 6. Low temperature C/T vs T data for annealed and unannealed A15 $Nb_3Si$, Recovery 20, expanded around 9 K to show the bcc Nb superconducting transition. The straight lines are guides to the eye to accentuate the transitions, marked by vertical lines. Based on published specific heat data for pure Nb[22], which give $\Delta C/T_c$ of 14 mJ/molK$^2$, the fraction of Nb in the Recovery 20 explosively compressed sample with 50-70 % A15 $Nb_3Si$ is 16%, while in the annealed sample the fraction of Nb has decreased to 5%. After the specific heat of the annealed sample was performed, it was ground and characterized by x-ray diffraction. Strong tetragonal $Nb_3Si$ diffraction lines were present.

### e.) Summary and Conclusions

A sample of A15 $Nb_3Si$ produced[6] by explosive compression (approximate peak pressure 90 – 110 GPa, peak temperature 1000 °C, residual temperature 500 °C) in 1981 at Los Alamos National Laboratory was subjected to pressures up to 88 GPa in order to determine the superconducting transition temperature $T_c$ as a function of pressure, with $T_c^{mid}$ falling to 5.2 K at the highest pressure. This sample was then annealed to approximately determine the temperature and time needed to transform A15 $Nb_3Si$ back into the tetragonal structure. Then, in order to determine if tetragonal $Nb_3Si$ can be transformed to the A15 structure at room temperature by applied pressures up to 92 GPa at room temperature (120 GPa at low temperatures), a sample of the original, before implosion LANL starting material (Nb$_3$Si annealed at 1780 °C for four hours, with ≥80% tetragonal structure) was measured.

Theoretical calculations of the enthalpies vs pressure of three possible Nb$_3$Si structures were carried out to help guide the pressure measurements. The result is that the A15 structure becomes the lowest enthalpy lattice arrangement above 50 GPa, with approximately 70 meV/atom lower enthalpy than the tetragonal structure at 100 GPa.

Although a second phase of Nb was present in all the samples, it was straightforward to distinguish which resistive transition corresponded to the A15 phase. No evidence of the A15 transition was detected in unloading data on the starting tetragonal material at 120, 105, and 83 GPa. Thus, we conclude that the original explosive compression conversion of tetragonal → A15 Nb$_3$Si at pressures up to 110 GPa indeed required the high temperatures (~1000 °C) associated with the rapid compression to assist the reaction kinetics. This is consistent with our annealing data on the A15, where temperatures of 600 °C for 5 days, as well as 4 days of lower temperature annealing, were required to convert 65% of the A15 phase (as determined by bulk specific heat measurements) back to normal Nb$_3$Si material.

Based on these experiments, and the original estimate[6] of the 1 μsec time-at-temperature (~1000 °C) and pressure (110 GPa) of the explosively compressed A15 Nb$_3$Si, the present work indicates that combined conditions of high pressure *and* high temperature are necessary to convert tetragonal Nb$_3$Si to the A15 structure. The long-lived nature of metastable A15 Nb$_3$Si (annealing for 39 years at room temperature failed to result in any sizable back transformation) points to the presence of a significant kinetic barrier. In that sense, it is not surprising that high temperatures are necessary to overcome this barrier. The present work highlights that high pressure phases formed with the addition of high temperatures are more likely to survive at ambient conditions than those formed via room temperature compression alone, since for the phase to form at room temperature the kinetic barrier must necessarily be small.

The barrier to atomic rearrangement raised by slow kinetics is the key to metastability at ambient conditions. The present work, where annealing at room temperature for 39 years did not transform the metastable A15 Nb$_3$Si structure back to the tetragonal (with its 90 meV/atom *lower* enthalpy) serves as a clear example of this. While computational methods now allow reliable prediction of thermodynamically stable phases, the prediction of such barriers and the associated kinetics are out of reach in most cases. Whether high temperature superconducting hydride phases can be retrieved to ambient conditions following ultra-high pressure synthesis thus remains an open question for experiments, though the high mobility of hydrogen may make this particularly challenging.

Given the current interest in the high-$T_c$ hydrides, the present work serves as a prototypical example of a superconducting structure that forms under very high (megabar) pressures but survives to ambient conditions.

**Acknowledgments**:  Work at the University of Florida performed under the auspices of US Department of Energy Basic Energy Sciences under Contract No. DE-SC-0020385.  Work at Los Alamos National Laboratory performed under the auspices of the U.S. Department of Energy under contract DE-AC52-06NA25396.